\documentclass[journal]{aiaa-pretty}

\usepackage{upgreek}
\usepackage{graphicx}
\usepackage{relsize}
\usepackage[capitalise]{cleveref}
\usepackage[version=4]{mhchem}
\usepackage{siunitx}
\usepackage{longtable}
\usepackage{adjustbox}
\usepackage{todonotes}

\author{Jan Rottmayer\thanks{Chair for Scientific Computing, University of Kaiserslautern-Landau (RPTU), Paul-Ehrlich-Straße 34, 67663 Kaiserslautern, Germany}, Emre Özkaya\thanksibid{1}, Sutharsan Satcunanathan\thanks{Institute of Aerodynamics and Chair of Fluid Mechanics, RWTH Aachen University, Wüllnerstraße 5a, 52062 Aachen, Germany}, Beckett Y. Zhou\thanks{Department of Aerospace Engineering, University of Bristol, University Walk, Bristol BS8 1TR, UK}, Max Aehle\thanksibid{1}, Nicolas R. Gauger\thanksibid{1},\\ Matthias Meinke\thanksibid{2}, Wolfgang Schröder\thanksibid{2}, Shaun Pullin\thanksibid{3}}

\title{Trailing-Edge Noise Reduction using Porous Treatment and Surrogate-based Global Optimization}

\date{\today}

\abstract{Broadband noise reduction is a significant problem in aerospace and industrial applications. Specifically, the noise generated from the trailing edge of an airfoil poses a challenging problem with various proposed solutions. This study investigates the porous trailing edge treatment. We use surrogate-based gradient-free optimization and an empirical noise model to efficiently explore the design space and find the optimal porosity distribution. As a result, a predicted $8-10$ dB reduction in the broadband $300-5000$ Hz was achieved. Furthermore, the optimal design emphasizes the design space's complexity and global exploration's difficulty. Further, the optimal design presents a low porous solution while constituting significant noise reduction.
}

\graphicspath{{./img}}
	
\begin{document}
\maketitle
\section{Nomenclature}
{\renewcommand\arraystretch{1.0}
\noindent\begin{tabular}{l c l}
	$\mathbf{a}$ & = & Parametrization vector $[a_1,a_2,a_3,a_4]$\\
	$b$ & = & Corcos's constant\\
	$c_0$ & = & Speed of sound \\
	$C_1, C_2, C_3$ & = & Goody's model constants \\
	$c_f$ & = & Forchheimer coefficient \\
	$d_p$ & = & Mean pore diameter \\
	$k$ & = & Wavenumber, $m^{-1}$ \\
	$k(\cdot, \cdot)$ & = & Correlation/Kernel function\\
	$\mathbf{k}$ & = & Correlation vector \\
	$\mathbf{K}$ & = & Correlation matrix\\
	$L$ & = & Airfoil span length\\
	$l_c$ & = & Chord length\\
	$l_y$ & = & Coherence spanwise length scale, m\\
	$M$ & = & Free-stream Mach number\\
	$m$ & = & Design dimension\\
	$n$ & = & Dimension of design space $\mathcal{X}$\\
	$P_{Ref}$ & = & Reference pressure \\
	$Re$ & = & Reynolds number \\
	$R_T$ & = & Ratio of timescales of pressure \\
	$S_0$ & = & Corrected distance to convection effects \\
	$S_{PP}$ & = & Far-field acoustic PSD \\
	$U_c$ & = & Convection velocity \\
	$U_e$ & = & External velocity $m \cdot s^{-1}$\\
	$U_i$ & = & Inlet velocity \\	
	$x, y$ & = & 2D spatial coordinates\\
	$\mathbf{x}$ & = & Design vector $[x_1,x_2,x_3,x_4,x_5]$\\
	$\alpha$ & = & Angle of attack \\
	$\beta_C$ & = & Clauser's parameter \\
	$\delta$ & = & Boundary-layer thickness \\
	$\delta^*$ & = & Boundary-layer displacement thickness \\
	$\boldsymbol{\theta}$ & = & Kernel parameters $[\theta_1, \dots, \theta_m]$\\
	$\Delta$ & = & Zadarola and Smits's parameter \\
	$\theta$ & = & Boundary-layer momentum thickness \\
	$\kappa$ & = & Permeability\\
	$\Pi$ & = & Wake strength parameter\\
	$\tau_w$ & = & Wall shear stress, Pa \\
	$\tau_{max}$ & = & Maximum shear stress, Pa \\
	$\phi$  & = &  Porosity\\
	$\Phi_{PP}$ & = & PSD of surface pressure fluctuations, $Pa^{2}/ Hz$\\
	$\mathcal{X}$ & = & Design space \\
	$\omega$ & = & Angular frequency \\	
	$\tilde{\omega}$ & = & Strouhal number based on external variables\\		
\end{tabular}}

\section{Introduction}
This paper presents a novel combination of existing methods for reducing broadband noise generated by an airfoil through the use of a porous trailing edge and surrogate-based global optimization. The trailing edge noise of an airfoil is the most challenging noise source to control in aerospace and industrial applications. The concept of this method is to create a porous segment on the trailing edge of the airfoil and then using numerical simulations to optimize the porosity distribution in order to minimize the broadband noise generated by the airfoil.\\
The optimization process is based on a surrogate model, a simplified mathematical representation of the complex physics governing noise generation. Using a surrogate model allows for efficient exploration of the design space and identification of the optimal porosity configuration of the porous trailing edge. The method does rely on the calculation of gradients, making them suitable for problems with complex or noisy objective functions.\\
The proposed method presents a combination of traditional computational fluid dynamics and computational aeroacoustics methods with an empirical surrogate for noise level prediction. This provides a new approach for optimizing the porous trailing edge in the context of noise reduction. \\
This study's results demonstrate the proposed method's effectiveness in reducing broadband noise and provide insight into the underlying physics of noise generation at porous trailing edges. The proposed method can be used as an additional tool in the noise reduction arsenal. It can be combined with other existing methods to achieve optimal noise reduction in aerospace and industrial applications such as wind turbines, aircraft, and industrial fans.

\section{Background}

\subsection{Design Space \& Parametrization}\label{sec:parametrization}
	The porosity model of the solver is a Darcy-Forchheimer porosity model described in~\cite{Koh2017J,satcunanathan2021,satcunanathan2022} and characterizes a porous material by its porosity $\phi$, permeability $\kappa$ and Forchheimer coefficient $c_f$. \\
	A challenging part of the design process is defining a suitable design space. The global optimization method's efficiency depends on the number of design dimensions. Therefore, a well-thought-out and informed parametrization will reduce the computational efforts necessary to find improved designs. \\
	For simplicity, we keep $c_f = 0.1$ constant. We take inspiration from the distribution investigated by~\cite{doi:10.2514/6.2015-2525}. To reproduce similar distributions and allow edge case scenarios, we define a function $f: [0,1] \times [-1, 1] \to [0, 1]$ as 
	\begin{equation}
		\begin{aligned}
			f(x, y) = a_1 x^{2 a_2} \hat{y}^{2 a_3}\\
			\hat{y} = \left(1 - a_4^\frac{1}{2 a_2}\right) \abs{y} + a_4^\frac{1}{2 a_2},
		\end{aligned}\label{eq:parametrization}
	\end{equation}
	where $a_1,a_2,a_3,a_4 \in [0,1]$ are the parameters.\\ 
	The function can be easily scaled to arbitrary bounds, and the variables $x, y$ can be interpreted as normalized coordinates. Furthermore, the function can easily be mapped to any quadrilateral element, such as the trapezoidal trailing edge investigated in this study. \\
	Note that for $a_2 = a_3 = a_4 = 0$, equation~\eqref{eq:parametrization} will result in a constant value defined by $a_1$ for all $x, y$. The parameters $a_2$ and $a_3$ control the dependence on the coordinate position, and if $a_2 \neq 0$ and $a_3 \neq 0$, $a_1$ defines the maximum value throughout the quadrilateral. $a_4$ was introduced to allow nonzero values on $y=0$ while not interfering with the coordinate dependence of $a_2$. In the given application to a trailing edge, $a_4$ can be interpreted as a chord condition. \\
	Equation~\eqref{eq:parametrization} was used to parameterize the material property \emph{porosity} $\phi$ in the trailing edge. Due to numerical problems with the porous-fluid interface computation in the solver, the function output was scaled between stable bounds $\phi_{\min} = 0.2$ and $\phi_{\max} = 0.9$
	\begin{equation}
		f^*(x, y) =  f(x,y) \cdot (\phi_{\max} - \phi_{\min}) + \phi_{\min}.
		\label{eq:scaling}
	\end{equation}\\
	As an additional parameter, we introduce the mean pore diameter $d_p \in [0.0001, 0.0035]$ to the design space. We utilize the material model used in~\cite{Koh2017J} and make use of the modified Ergun equation~\cite{Whitaker1996} to compute the permeability $\kappa$ as a function of porosity $\phi$ and mean pore diameter $d_p$. The bounds on $d_p$ are chosen so that the resulting permeability closely matches the experimental results of~\cite{Alomair2018ExperimentalMO} for open foam aluminum samples. \\
	The resulting design space is $\mathcal{X}: [0,1]^4 \times [0.0001, 0.0035]$.
	
\subsection{Efficient Global Optimization}\label{ss:EGO}
 	The Efficient Global Optimization (EGO) algorithm, first proposed by~\cite{ego}, presents an optimization scheme for expensive black-box functions without prior knowledge about the response function $f(\mathbf{x}): \mathbb{R}^n \to \mathbb{R}$. The core idea is the reduction of computational complexity by approximating the response function with a surrogate and exploiting the surrogate to find sample locations most likely to show improvement. We will cover a short summary of the algorithm. For more details refer to~\cite{ego}. The schematic of the algorithm is shown in figure~\ref{fig:ego}. 
 	
 	\subsubsection{Design of Experiment}
\begin{figure}
		\includegraphics[width=1.0\linewidth]{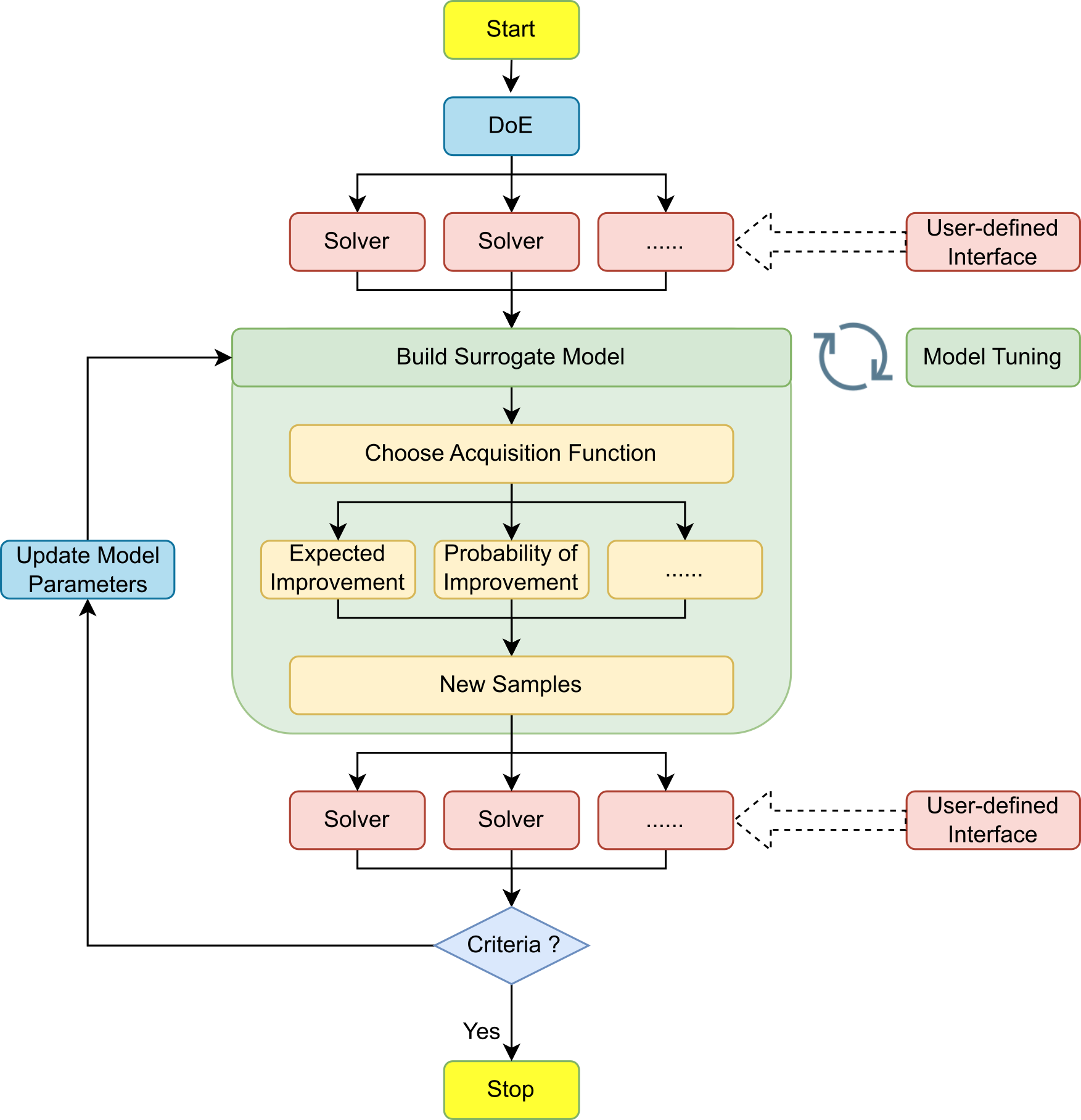}
		\caption{Schematic of the EGO framework. The solver block is understood as union of the computation chain from design based initialization in the RANS solver, the RANS solver, extraction of boundary layer variables, to the computation of the objective from the trailing edge noise model~\ref{ss:TEN}.}
		\label{fig:ego}
	\end{figure}

 	The initial problem is gathering information about the response function optimally, referred to as ``Design of Experiment`` (DoE). We use Latin Hypercube combined with the minimax design criterion formulated in~\cite{JOHNSON1990131}. This combination is well known in literature and is usually referred to as optimal Latin Hypercube design. \\
 	Let $\mathcal{D}$ denote the design space of all possible Latin Hypercube designs, $d(\cdot, \cdot)$ be the Euclidean distance function defined on $\mathcal{D} \times \mathcal{D}$. Then, for $\mathcal{X} \subset \mathcal{D}$ with $\#\mathcal{X} = n$, $\mathcal{X}^*$ is the set of designs optimizing the space-filling condition 
 	\begin{equation}
 		\min_{\mathcal{X}} \max_{\mathbf{x} \in \mathcal{X}}  d(\mathbf{x}, \mathcal{X}) = \max_{\mathbf{x} \in \mathcal{X}} d(\mathbf{x}, \mathcal{X}^*). 
 	\end{equation}
	The designs $\mathbf{x}_i \in \mathcal{X}^*$ are evaluated with the response function to get $\mathbf{y} = [y_1, \dots, y_n]^T$. The evaluation with the response function is denoted as \emph{Solver} in figure~\ref{fig:ego}.

	\subsubsection{Surrogate Model}
	A stochastic process model is trained, known as Gaussian process regression with the Kriging kernel~\cite{doi:https://doi.org/10.1002/9780470770801}.
	The statistical process model results in a point-predictive distribution
	\begin{equation}
		p(y \vert \mathbf{x}, \mathcal{X})~ \mathcal{N}(y \vert \mu(\mathbf{x}), \sigma^2(\mathbf{x}))
		\label{eq:gprregression}
	\end{equation}
	with the mean $\mu(\mathbf{x})$ and standard deviation $\sigma(\mathbf{x})$. Usually, $\mu(\mathbf{x})$ is defined as a fixed polynomial function. \\
	
	Let $\mathbf{K}$ denote the correlation matrix with entries $\mathbf{K}_{i,j} = k(\mathbf{x}_i, \mathbf{x}_j)$ and $k(\cdot, \cdot)$ denote a kernel function, e.g., the multivariate Gaussian kernel 
	\begin{equation*}
		k(\mathbf{x}_i, \mathbf{x}_j) = \exp{\left(- \sum_{l=1}^{m} \frac{1}{\theta_l}\left(\mathbf{x}_i^{(l)} - \mathbf{x}_j^{(l)}\right)^2\right)},
	\end{equation*}
	where $\mathbf{x}_i^{(l)}$ denotes the $l$-th entry of the $i$-th sample in $\mathcal{X}$ and $\boldsymbol{\theta} = \{ \theta_1, \dots, \theta_m\}$ are hyperparameters.
	Further, $\mathbf{k}$ is the correlation vector defined by $\mathbf{k}(\mathbf{x}) = [k(\mathbf{x}, \mathbf{x}_1), \dots, k(\mathbf{x}, \mathbf{x}_n)]^T$ and $\boldsymbol{\mu} = [\mu(\mathbf{x}_1), \dots, \mu(\mathbf{x}_n)]^T$ $\forall \mathbf{x}_i \in \mathcal{X}$ is the vector of predictions with the polynomial. Then 
	\begin{equation}
		\hat{\beta} = \left(\boldsymbol{\mu}^T \mathbf{K}^{-1} \boldsymbol{\mu}\right)^{-1} \boldsymbol{\mu}^T \mathbf{K}^{-1} \boldsymbol{y}
	\end{equation}
	defines the maximum likelihood (MLE) Kriging predictor
	\begin{equation}
		\hat{f}(\mathbf{x}) = \hat{\beta} + \mathbf{k}^T(\mathbf{x}) \mathbf{K}^{-1} (\mathbf{y} - \boldsymbol{\mu} \hat{\beta}).\label{eq:KrigPred}
	\end{equation}
	The estimate of the variance is
	\begin{equation}
		\sigma^2 = \frac{(\mathbf{y} - \boldsymbol{\mu} \hat{\beta})^T \mathbf{K}^{-1}(\mathbf{y} - \boldsymbol{\mu} \hat{\beta})}{n}
	\end{equation}
	For the details and the derivation of the Kriging predictor in equation~\eqref{eq:KrigPred}, refer to~\cite{doi:https://doi.org/10.1002/9780470770801}.
	
	\subsubsection{Acquisition Function}
	We use the expected improvement (EI) as an acquisition function. There are many other alternatives, as shown by~\cite{frazier2018}, but we will focus on EI alone.\\
	Let $I(\mathbf{x})$ denote the improvement function 
	\begin{equation}
		I(\mathbf{x}) = \begin{cases}
			y(\mathbf{x}) - y_{\min}, &\text{ if } y(\mathbf{x})<y_{\min}\\
			0, &\text{else},
		\end{cases}
		\label{eq:improvement}
	\end{equation} 
	where $y_{\min}$ is the minimal observed function value from the samples in $\mathcal{X}$. Then the expected improvement is
	\begin{equation}
		\mathbb{E}\left[I(\mathbf{x})\right] = \int_{-\infty}^{\infty} I(\mathbf{x}) p(z) dz,\label{eq:EI}
	\end{equation}
	where $p(z)$ is following from equation~\eqref{eq:gprregression}.\\
	In EGO, equation~\eqref{eq:EI} is minimized to determine the best sample locations for optimization of $f(\mathbf{x})$. Note that the expected improvement considers the amplitude and probability of a possible improvement. Both considerations are only possible due to the predictions provided by the Kriging predictor. 

\subsection{Amiet-based Trailing Edge Noise Model\label{ss:TEN}}
For the prediction of the Sound Pressure Spectrum (SPL) of an angular frequency $\omega$ at a location $\mathbf{x}$ we use an extension to the Amiet's trailing-edge noise model proposed in~\cite{ROGER2005477} 
\begin{equation}
	S_{PP}(\mathbf{x}, \omega) = \left(\frac{\omega b x_3}{2 \pi c_0 S_0^2}\right)^2 \frac{L}{2}\left\vert I\left( \frac{\omega}{U_c}, \frac{k x_2}{S_0}\right) \right\vert l_y(\omega) \Phi_{PP}(\omega)	\label{eq:ATENModel}
\end{equation}
by considering the adverse pressure gradient effects in the empirical wall pressure spectral model
\begin{equation}
	\Phi_{PP}(\omega) = \frac{\tau_{max}^2 \delta ^*}{U_e} \frac{\left[ 2.82 \Delta^2(6.13\Delta^{-0.75} + F_1)^{A_1}\right] \left[4.2\left(\frac{\Pi}{\Delta}\right)+1\right] \tilde{\omega}^2}{\left[4.76\tilde{\omega}^{0.75}+F_1\right]^{A_1} + \left[C_3^{'} \tilde{\omega}\right]^{A_2}},\label{eq:WPSModel}
\end{equation}
proposed by~\cite{doi:10.2514/1.J051500}. The parameters are given as
\begin{align*}
	\tilde{\omega} &= \frac{\omega\delta^*}{U_e}\\
	A_1 &= 3.7 + 1.5 \beta_C\\
	A_2 &= \min\left(3, \frac{19}{\sqrt{R_T}}\right) + 7\\
	F_1 &= 4.76 \left(\left(\frac{1.4}{\Delta}\right)^{0.75} \left[0.375 A_1 - 1\right]\right)\\
	\beta_C &= \frac{\theta}{\tau_w}\frac{dp}{dx}\\
	R_T &= \frac{\delta U_e}{\nu u_\tau^2}\\
	\Pi &= 0.8 (\beta_C + 0.5)^\frac{3}{4}\\
	\Delta &= \frac{\delta}{\delta^*}\\
\end{align*}
For the details, refer to~\cite{ROGER2005477} and~\cite{doi:10.2514/1.J051500}. \\
We compute the SPL as 
\begin{equation}
	\text{SPL}(S_{PP}) = 10 \log\left(\frac{2 \pi S_{PP}}{P_{Ref}^2}\right)\label{eq:SPL},
\end{equation}
from the $S_{PP}$ predicted by equation \eqref{eq:ATENModel}. We can think of equation~\eqref{eq:SPL} as function of boundary layer parameters, which can be extracted directly from the RANS solutions provided by the solver. It should be made clear here, the WPS model prescribed in equation~\eqref{eq:WPSModel} is a surrogate model and validated on some sample data from several experiments. 

\subsection{Broadband Noise Reduction}
The objective is the reduction of broadband noise generated from the trailing edge by non-homogeneous porous treatment. We consider frequencies $\omega$ in the range of 300 - 5000Hz. Further, we consider the SPL at different spacial locations defined by a circle with a radius of $1.5 l_c$ around the trailing edge. We define the broadband noise objective as minimization problem
\begin{equation}
\begin{aligned}
	\min_{\mathbf{x} \in \mathcal{X}} & \int_{0}^{2\pi}\int_{\omega_{\min}}^{\omega_{\max}}  \text{SPL}(\mathbf{x}, \omega, \theta) d \omega d\theta, 
	\label{eq:objective}
\end{aligned} 
\end{equation}
where $\theta$ is the polar angle and $\mathcal{X}$ is the space of possible designs. Note no constraints are set concerning aerodynamic performance. Equation~\eqref{eq:objective} can be seen as function of the design $\mathbf{x}$ alone. Thus, we can use the black-box optimization framework provided by EGO, as described in section~\ref{ss:EGO}. 

\section{Computational Setup}		
In the problem setup, we consider a NACA 0012 aerofoil with a blunt trailing edge at a Mach number of $M=0.1118$, a Reynolds number $Re=1e6$, and an angle of attack of $\alpha = 0$. The RANS equations are solved on a two-dimensional structured multi-block mesh with a porous trailing edge of length $0.15 l_c$ and compared to the equivalent setup for a solid trailing edge configuration. The RANS solver is based on the ansatz of volume-averaging before Reynolds averaging~\cite{MONER201525} and takes the additional terms into account that arise from the extension to porous materials as described in~\cite{MONER201525,satcunanathan2022}. The porosity and permeability distributions are initialized based on the designs described in section~\ref{sec:parametrization}. Boundary layer parameters are extracted at the trailing edge solution. The wall pressure spectral model in equation~\eqref{eq:WPSModel} is evaluated, and noise level predictions are computed with the Amiet-based trailing edge noise model of equation~\eqref{eq:ATENModel} and equation~\eqref{eq:SPL}. The objective is the minimization problem stated in equation~\eqref{eq:objective} and calculated with a trapezoidal integration rule to complete the computation chain. The union of methods constitutes the black-box function $f(\mathbf{x}): \mathcal{X} \to \mathbb{R}$ that we want to minimize with the EGO algorithm, as described in section~\ref{ss:EGO}. The RANS solver uses a time marching scheme; thus, we used 3000000 time steps to assure convergence. To further compare our results, we simulate a homogeneous porous trailing edge with $\phi = 0.305 = const.$, which is the corresponding maximum porosity of the optimal design.  
	
\section{Results}
\begin{figure}
	\includegraphics[width=0.8\linewidth]{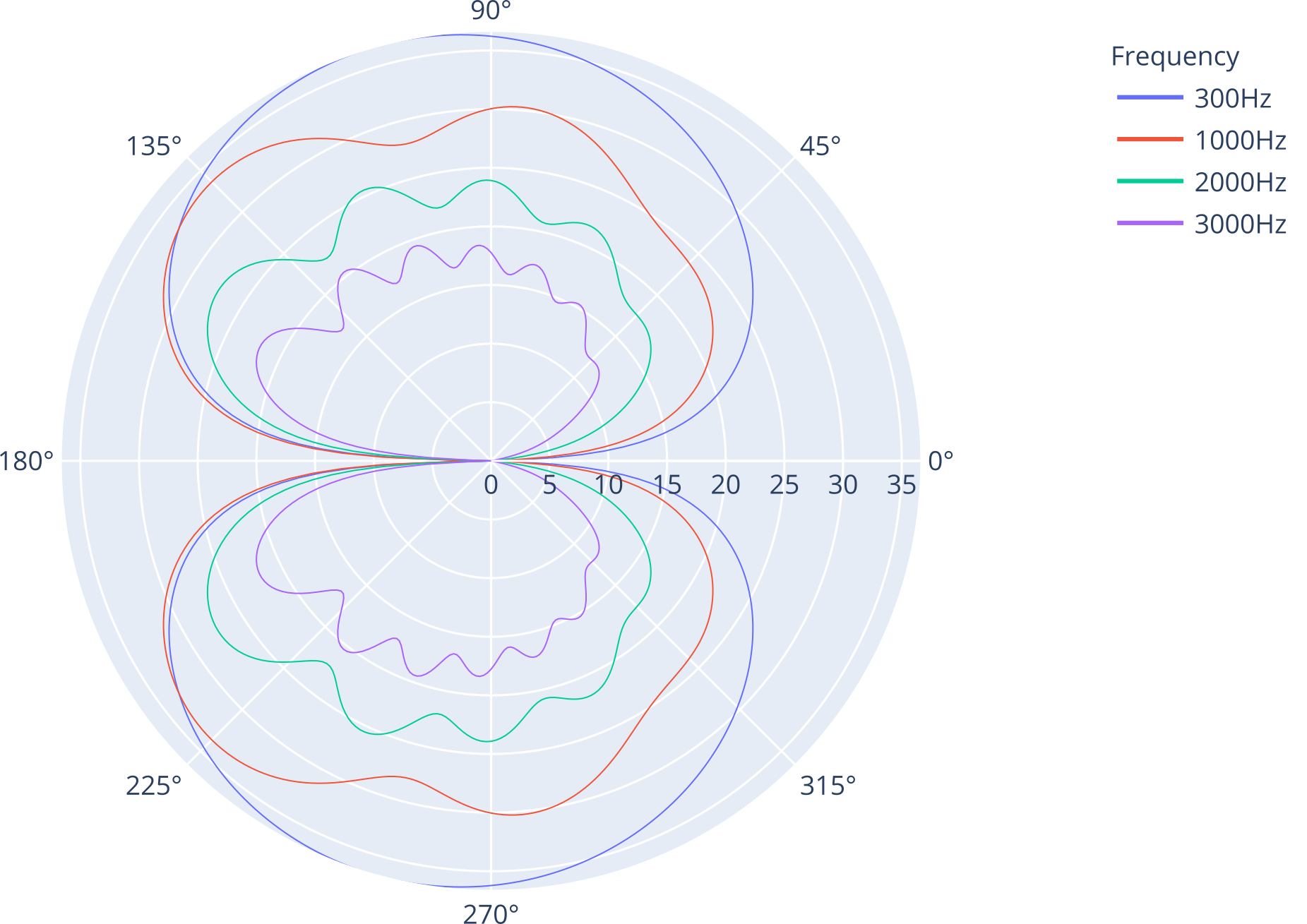}
	\caption{SPL prediction of different frequencies depending on polar angle at constant radius $R=1.5 l_c$ for a solid trailing edge configuration with objective value $0.493\cdot 10^{6}$.}
	\label{fig:solidSPL}
\end{figure}
\begin{figure}
	\includegraphics[width=0.8\linewidth]{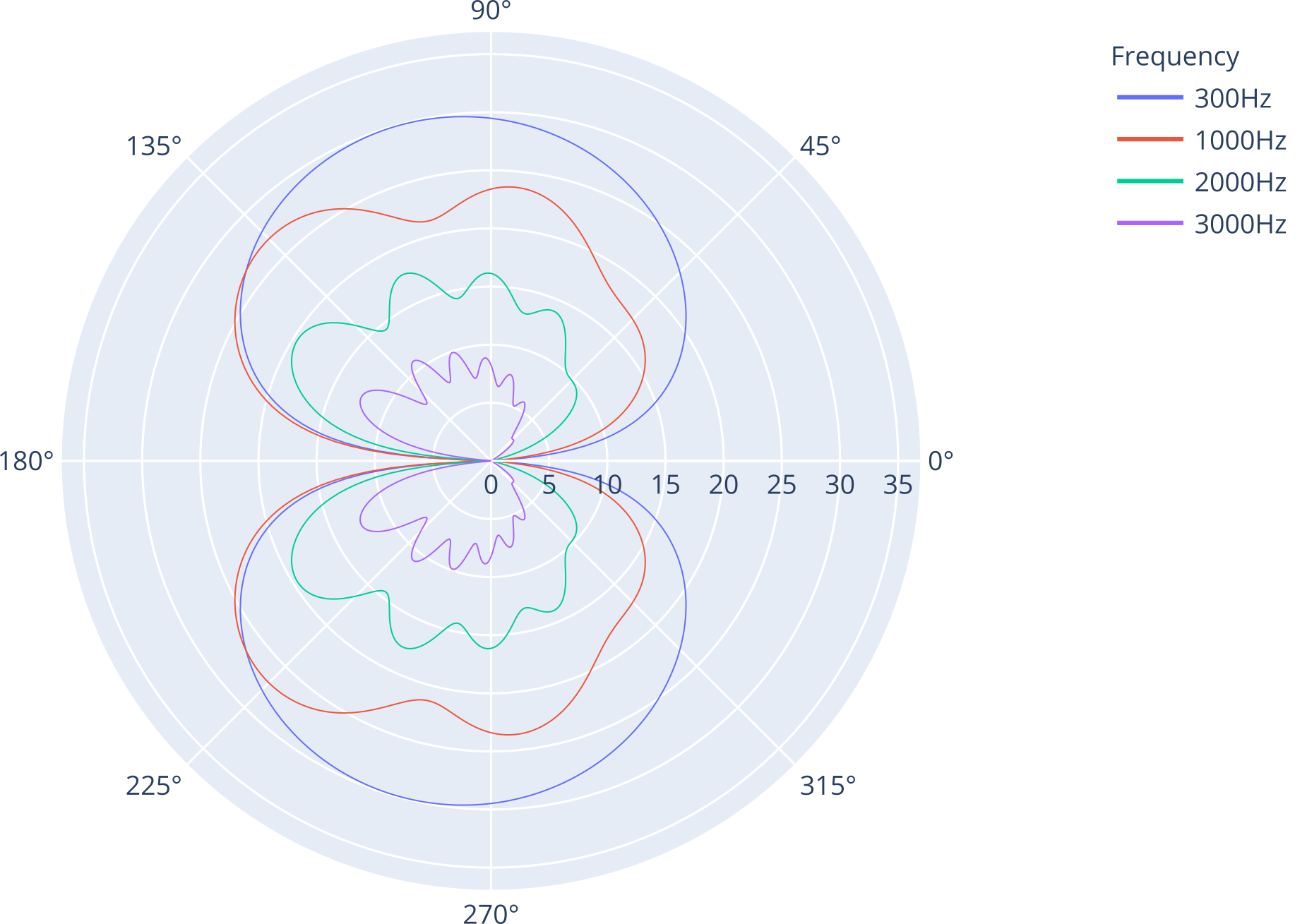}
	\caption{SPL prediction of different frequencies depending on polar angle at constant radius $R=1.5 l_c$ for the optimal porous trailing edge configuration $\mathbf{x}_{opt} = [0.15, 0.11, 0.77, 0.77, 0.0001]$ with objective value $0.269 \cdot 10^{6}$.}
	\label{fig:optimalSPL}
\end{figure}
\begin{figure}
	\includegraphics[width=\linewidth]{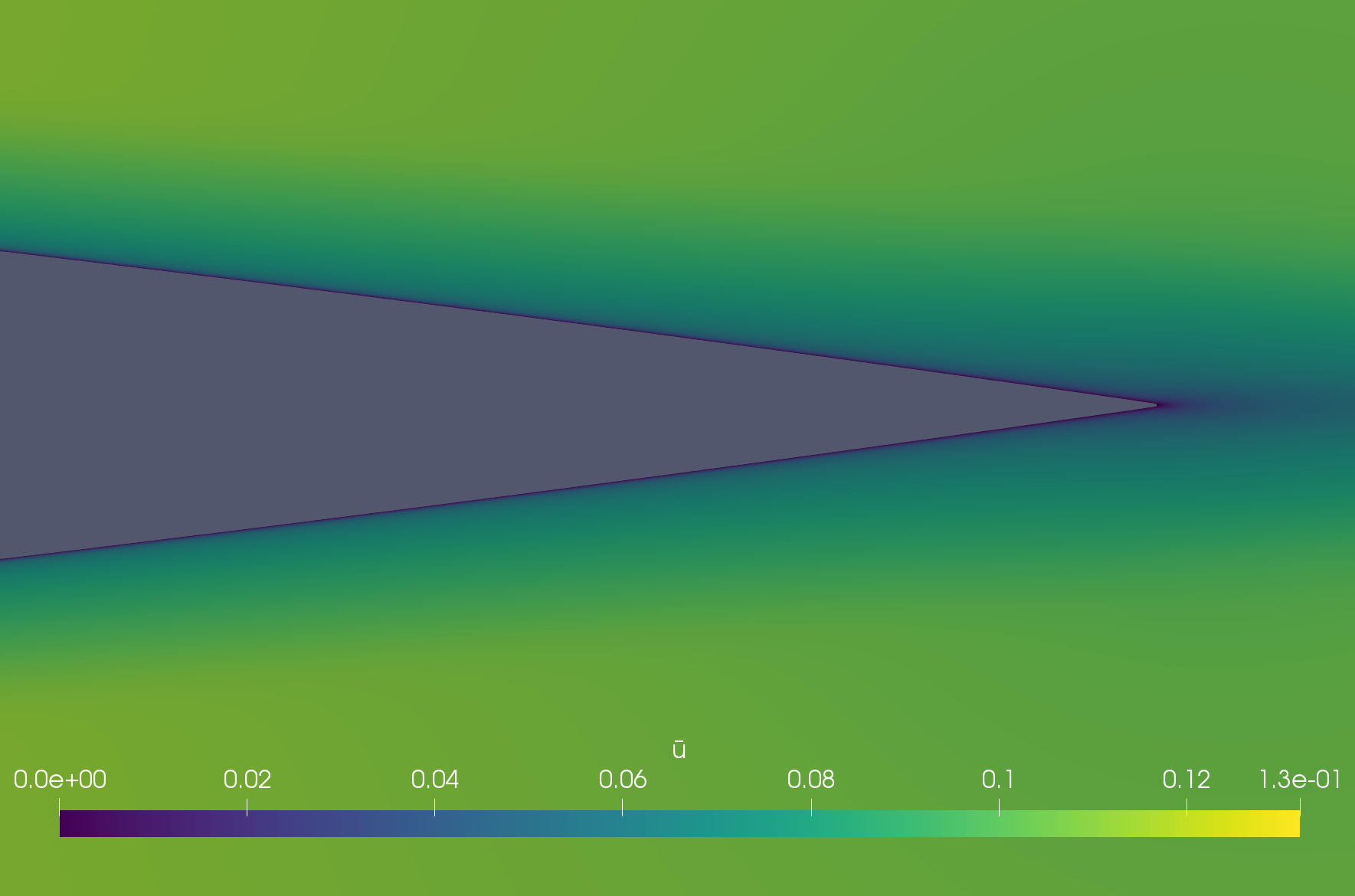}
	\caption{Mean velocity component $\bar{u}$ at the trailing edge for the solid configuration. }
	\label{fig:solidu}
\end{figure}
\begin{figure}
\end{figure}
\begin{figure}
	\includegraphics[width=\linewidth]{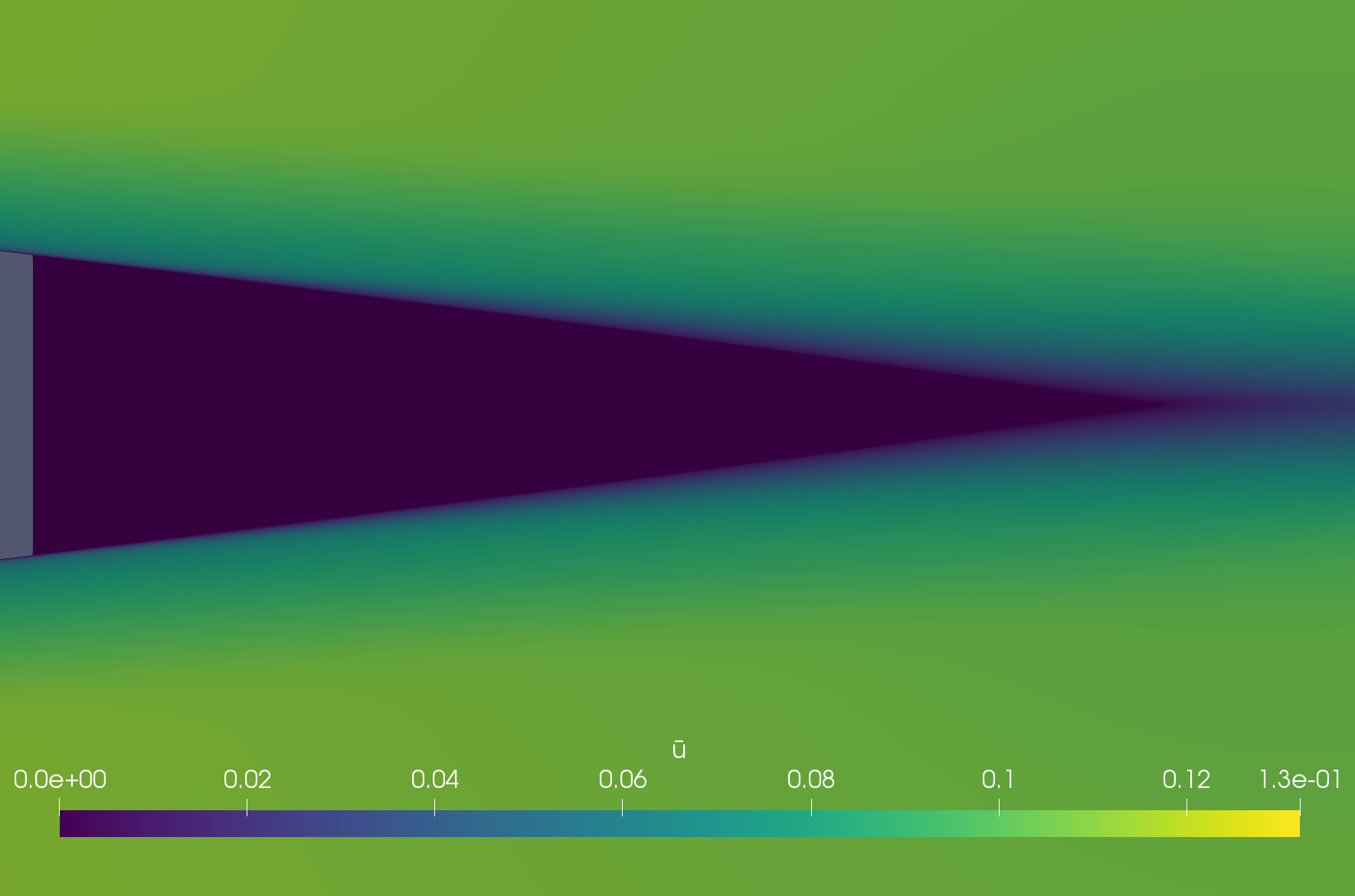}
	\caption{Mean velocity component $\bar{u}$ at the trailing edge for optimal porous treatment. }
	\label{fig:optimalu}
\end{figure}

\begin{figure}
 	\includegraphics[width=\linewidth]{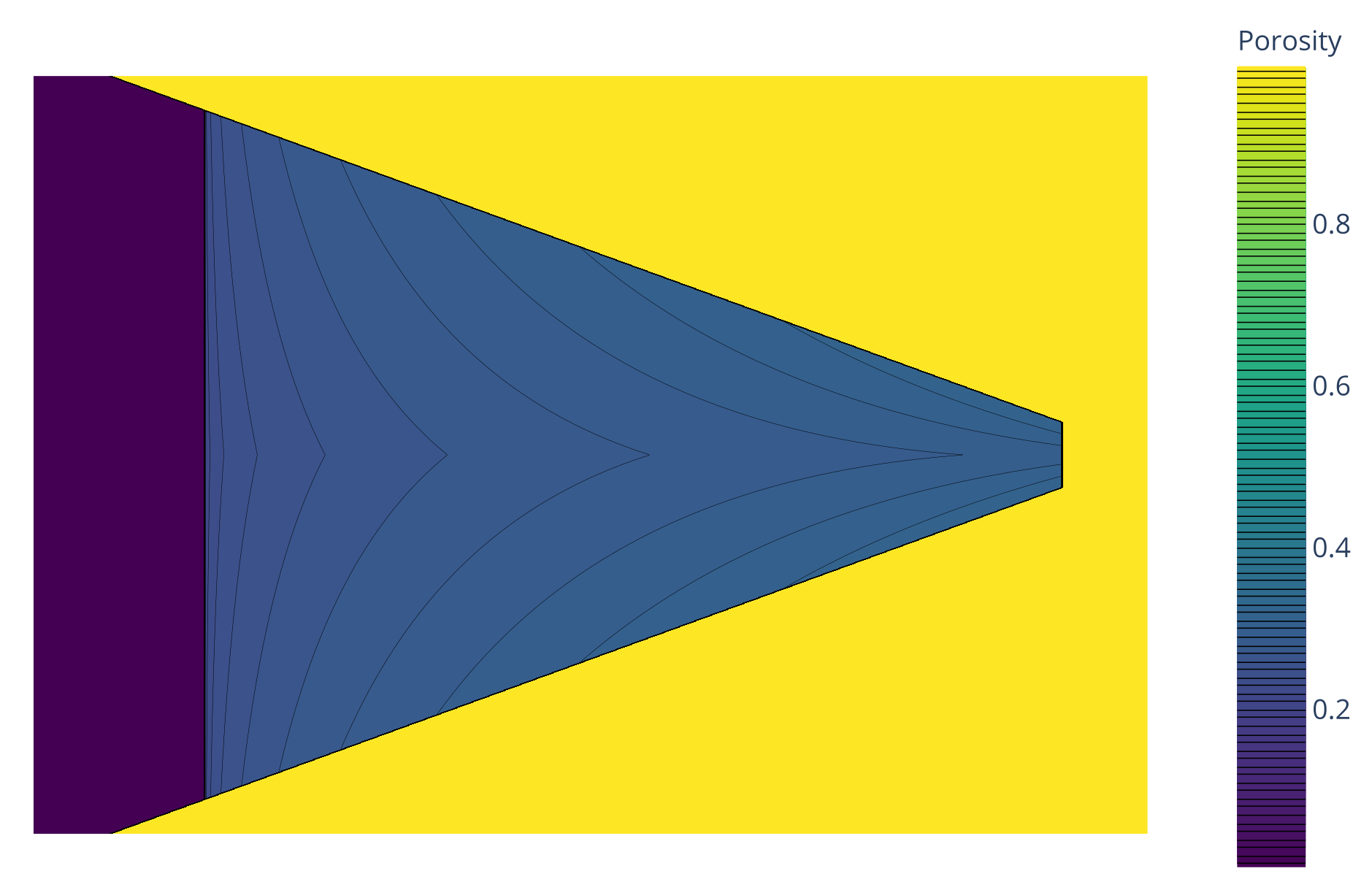}	
 	\caption{Porosity distribution in trailing edge defined by $\mathbf{x}_{opt} = [0.15, 0.11, 0.77, 0.77, 0.0001]$. The trailing edge model distorts width and height for clarification.}
 	\label{fig:optimalDesign}
\end{figure}

The initial design of experiment phase already highlights the variation and complexity of the response surface. The objective values range from $0.287\cdot 10^{6}$ to $0.442\cdot 10^{6}$. The solid trailing edge produces an objective value of $0.493\cdot 10^{6}$. \\
The SPL of the default design is shown in figure~\ref{fig:solidSPL}, and the corresponding mean flow component $\bar{u}$ in figure~\ref{fig:solidu}. The results of the best sample after $20$ EGO iterations are shown in figures~\ref{fig:optimalSPL} and~\ref{fig:optimalu}. The EGO algorithm has yet to explore much of the design space. The best sample improved to $0.269\cdot 10^{6}$ for the design $\mathbf{x}_{opt} = [0.15,0.11,0.77,0.77,0.0001]$. In the case of the homogeneous configuration, an improvement to $0.257\cdot 10^{6}$ was achieved. \\
A general observation of the SPL, the amplitude decrease with increasing frequency, and there appears to be a frequency-dependent oscillation pattern in the amplitude. \\ 
The initial Design of Experiment (DoE) resulted in multiple regions with low functional value. The most promising samples taken from the DoE are given in~\ref{tb:promising_samples}.
\begin{table}
	\centering
	\begin{tabular}[h!]{l |c }
		\hline
		$\mathbf{x}$ & $f(\mathbf{x})$ \\
		\hline
		\hline
		$[0.12,0.60,0.91,0.17,0.0001]$ & $0.3\cdot 10^{6}$\\
		$[0.83,0.71,0.83,0.33,0.00017]$ & $0.287\cdot 10^{6}$\\
		$[0.33, 0.25, 0.45, 0.43, 0.00024]$ & $0.288\cdot 10^{6}$\\
	\end{tabular}
\caption{Most promising design regions from the initial design of experiment exploration of the design space.}
\label{tb:promising_samples}
\end{table}
An $8-10$ dB decrease in SPL was achieved by the optimal design of the EGO. The comparison of the flow field component $\bar{u}$ between solid and optimal design in figures~\ref{fig:solidu} and~\ref{fig:optimalu} show a significantly widened boundary layer for the optimized design. Note that the pressure drop across the trailing edge is significantly reduced for the optimal design. The porosity distribution of the optimal design is visualized in figure~\ref{fig:optimalDesign}. 
 
\section{Discussion \& Conclusion}
Our findings of a potential improvement of $8-10$ dB in the broadband range $300-5000$Hz align well with the investigations of~\cite{doi:10.2514/6.2015-2525}. The relationship between design space and response is highly complex, induced by the fluctuations observed in the design samples. The improvement to $0.269\cdot 10^{6}$ by $\mathbf{x}_{opt} = [0.15, 0.11, 0.77, 0.77, 0.0001]$ is an exciting finding. The resulting flow solution in figure~\ref{fig:solidu} gives a small insight into a possible explanation. The boundary layer thickness is increased, indicating an interaction between the flow and the porous material. Although the flow is primarily similar, a significant decrease in the magnitude of the pressure gradient across the trailing edge is present. The pressure fluctuation across the trailing edge is a crucial mechanism for noise generation. The porous trailing edge allows a more gradual pressure equalization.\\
This probably results in a decrease in aerodynamic performance, although not quantified in this study. Further, we should make reserved conclusions due to the fact that we are only optimizing an empirical approximation of trailing edge noise. Our minimizer might just be a design for which the noise level is strongly underestimated instead of a genuine improvement. Since the resulting prediction of noise levels closely aligns with the findings in~\cite{Koh2017J} and does not contradict the noise levels found at $M=0.2$ in~\cite{doi:10.2514/1.J051500}, we assume the trailing edge noise model to predict the SPL spectrum correctly. \\
Therefore, we attribute the broadband reduction of SPL to the dampening effect of the porous material. The non-homogeneous aspect of the material properties results in an additional benefit not captured by the noise model. The porous solid interface poses an abrupt change in flow resistance and thus will produce pressure fluctuations and thus noise. This additional noise source is completely ignored in the applied model. This explains the additional improvement found by the homogeneous configuration. A trivial optimization solution in this setup is a fully porous configuration, essentially pushing the true interactions away from the sensor location and thus invalidating the model's predictions. \\
The comparison to a homogeneous porous trailing edge at the optimal porosity, defined by $x_1$ of the optimal design $\mathbf{x}_{opt}$, shows an additional improvement of the objective to $0.257\cdot 10^{6}$, which does not take into the additional noise generated from the abrupt change. The change in the SPL amplitudes would be indistinguishable from the SPL spectrum shown in figure~\ref{fig:optimalSPL}. Some investigations have shown that for $d_p = 0.0001$ the best results are achieved. This collapses the design dimension to $4$ and initially makes the algorithm less efficient. Due to the limited sampling of the design space and the limited number of optimization iterations, the achieved optimal presents only a local minimum. By increasing the number of iterations, the optimizer should converge to a better non-homogeneous configuration due to an increase in the approximation's quality. Any of the previously mentioned problems can disrupt the convergences. \\ 
The improvement shows that EGO can significantly reduce noise in a low number of optimization steps. The magnitude of improvement is similar to the non-homogeneous configurations investigated in~\cite{doi:10.2514/6.2015-2525, Herr2007,Sarradj2007NoiseGB, Koh2017J}. Therefore, we consider our findings cautiously to be in line with the literature. \\
The interpretation of the optimal design variables allows some assumptions to be drawn. The low value of $x_2$ expresses the low dependence of the position along the chord. The high values of $x_3$ and $x_4$ indicate the importance of the depth of the material. The primary interaction with the fluid happens at the surface level due to the flow resistance of the material defined by the permeability $\kappa$. Lastly, the value of $x_1$ hints that we do not need highly porous materials to achieve a significant change in the produced noise. \\  

A problem with the trailing edge noise computation chain is the evaluation of boundary layer parameters on a single location, namely the far end of the trailing edge. This opens up space for problems due to numerical oscillation and unresolved mean flow. The RANS solver uses a time marching scheme and shows different convergence speeds depending on the porosity configuration. This could partly be seen in the solution by having laminar vortex shedding present. Therefore, the number of time steps was purposefully set much higher than what is typically required for convergence. Additionally, a trivial solution arises, being a fully porous trailing edge. The noise model setup cannot capture the additional noise generated from the transition from solid to porous. \\
Further, the employed wall pressure spectral model in equation~\eqref{eq:WPSModel} still needs to be validated for porous surface applications. The trailing edge noise model in equation~\eqref{eq:ATENModel} solely dependents on the boundary layer parameters and states an empirically derived model from similar airfoil configurations and flow conditions. Therefore, the quality of the prediction capabilities is investigated in~\cite{ROGER2005477} and~\cite{doi:10.2514/1.J051500}. \\
Lastly, the definition of broadband noise in equation~\eqref{eq:objective} takes a uniformed ansatz. The formulation of the objective presents a uniformly weighted function concerning the angular location and the frequency. Although more sophisticated weights might be helpful, the chosen formulation is sufficiently targeted for general broadband noise reduction.

A significant part of outstanding work is the validation of the different computation modules. Specifically, the validation of the wall pressure spectral model, used in the noise computation, with some computational aeroacoustics method, for example, the one explored in~\cite{Koh2017J}, is required to apply to porous materials flows. An initial validation of the optimal design should confirm the predicted noise reduction. To generalize the optimal design, the consideration of aerodynamic performance and different flow conditions, e.g., the variation to different angles-of-attack $\alpha$, is necessary. For a full conclusion on the optimality of the designs, the design space should be explored more rigorously, and the noise generated by the solid porous transition should be considered.

\section{Acknowledgment}
The support of this research by the Deutsche Forschungsgemeinschaft DFG under grant GA 857/6-2 is gratefully acknowledged. 

\section{Summary}
We have shown the ansatz of non-homogeneous porous treatment of the trailing edge is superior to the homogeneous treatment. We combined a RANS solver, an empirical noise model and a global optimization method to minimize broadband noise for an NACA 0012 airfoil at a single flow configuration. We predicted a reduction of noise levels by $8-10$ dB across all frequencies (reference maximum is $38$ dB) with a minimally porous treatment.     
 
\nocite{*}
\bibliographystyle{amsplain}
\bibliography{bibliography}

\end{document}